\documentstyle[11pt,newpasp,twoside,epsf]{article} 
\def\edcomment#1{\iffalse\marginpar{\raggedright\sl#1\/}\else\relax\fi}
\reversemarginpar

\begin{document}
\title{The Colour-Magnitude Relation and the Age of Galaxies in the Cluster AC118 at 
z=0.31}
\author{P. Merluzzi, G. Busarello, M. Massarotti  }
\affil{Osservatorio Astronomico di Capodimonte, Napoli, Italy}
\author{F. La Barbera}
\affil{Physics Department, Universit\`a Federico II, Napoli, Italy}

\begin{abstract}
We derive the Colour-Magnitude (CM) relation for the galaxy cluster
AC118 at z=0.31 and use this relation to set constraints on the formation
epoch and on the age spread of cluster galaxies in the framework
of the passive evolution scenario. This work is based on observations 
at ESO NTT (OAC guaranteed time).
\end{abstract}

The galaxy population in the core of AC118 was already analyzed by
Stanford et al. (1998), who found evidence in favour of the passive 
evolution scenario. Barger et al. (1996), on the other hand, claimed
for evidence of recent bursts of star formation.
Here we derive the CM relation in the optical and near infrared (R-K) for a 
large sample of cluster
members (N$_{gal}$=318) on a wide area ($\sim$2.6 Mpc$^2$, q$_0$=0.5,
H$_0$=50 Km s$^{-1}$Mpc$^{-1}$) around the cluster core.

\begin{figure}[!h]
\plotone{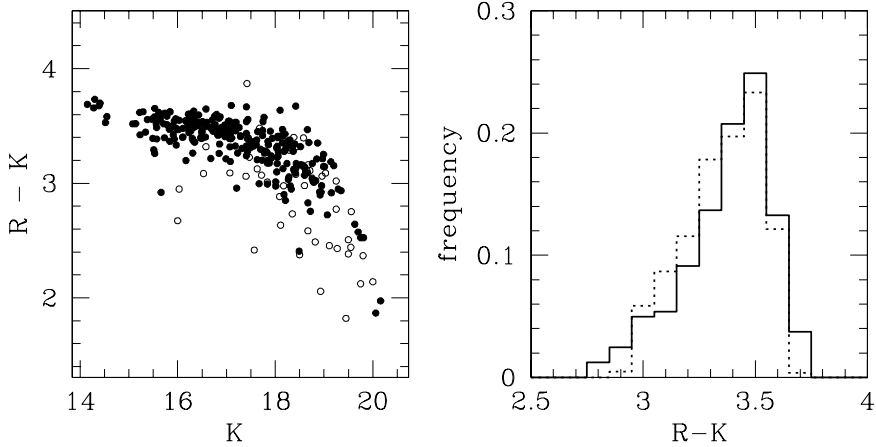}
\caption{{\it Left panel}: CM relation of the cluster members
selected through photometric redshifts. Filled/open dots mark galaxies
described by early--/late--type templates. Typical uncertainties are 
$\Delta$(R-K)=0.04 mag and $\Delta$K=0.02 mag. {\it Right panel}: observed R-K
colour distribution (solid line) compared with the distribution
predicted by the adopted model of galaxy evolution 
(dotted line).}
\end{figure}

The sample was selected by a photometric redshift technique which
is based on the Spectral
Energy Distribution method (e.g. Massarotti, Iovino, \& Buzzoni 2001)
using the template spectra provided by the code by Bruzual \& Charlot
(1993).

\begin{figure}
\plotone{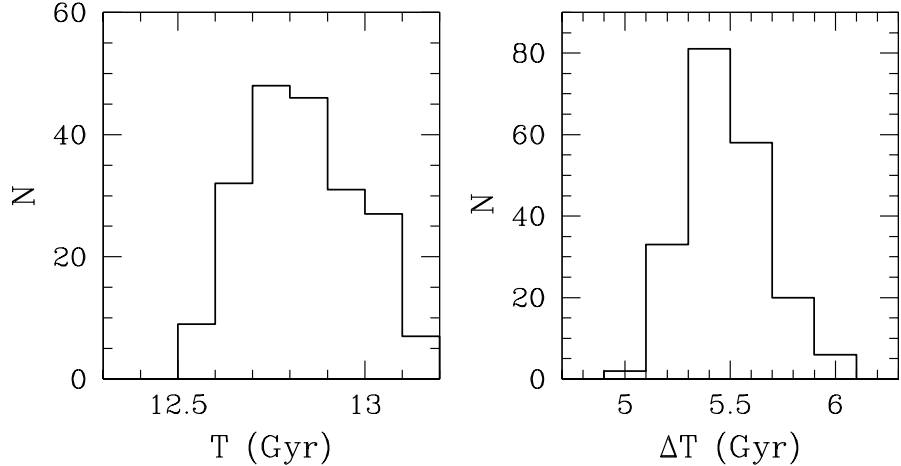}
\caption{{\it Left panel}: distribution of the first epoch of star
formation as obtained by N=200 sets of simulations of the R-K galaxy
colours.  {\it Right panel}: distribution of the age spread.}
\end{figure}

Figure 1 (left panel) shows the CM relation of the selected
cluster members. We analyzed the R-K colour distribution
of the galaxies that are described by early--type templates. 
The observed distribution (solid line in Figure 1) can be
recovered by a simple model of galaxy formation (dotted line): stars
in galaxies form in short bursts ($\tau$=1 Gyr) randomly occurring in
a time interval ($T-\Delta T$, $T$). $T$ and $\Delta T$ are the first
epoch of star formation and its duration. For given values of
$T$ and $\Delta T$, the expected R-K colour distribution was
simulated taking into account the measurement uncertainties on the
colours. The best fit estimates of $T$ and $\Delta T$ were derived from
different sets of simulations. In Figure 2 (left panel) we show the 
distribution of the first epoch of star formation. In the
adopted cosmological model galaxies form the oldest stars very early
in the past ($z>7$).

The spread in star formation times is shown in the right panel of Figure 2. 
To describe the colours of the bluest cluster galaxies, these
objects should have undertaken a recent ($T<$8 Gyr, $z<$1)  burst of star formation.


\begin{references}

\reference Barger, A.J., Arag\'on--Salamanca, A., Ellis, R. et al.  1996, \mnras, 279, 1

\reference Bruzual, G.A., \& Charlot, S. 1993, \apj, 405, 538

\reference Massarotti, M., Iovino, A., \& Buzzoni, A. 2001, \aap, 368, 74

\reference Stanford, S.A., Eisenhardt, P.R.M., \& Dickinson, M. 1998, \apj, 
492, 461

\end{references}
\end{document}